\documentclass[aps,prb,preprint,preprintnumbers,amsmath,amssymb,floatfix,showpacs]{revtex4}
\usepackage{graphicx}
\usepackage{dcolumn}
\usepackage{bm}
\usepackage{color}

\begin{document}
\preprint{Phys. Rev. B}

\title{Assessment of the Thermal Conductivity of BN-C Nanostructures}
\author{Alper K{\i}nac{\i}}
\affiliation{%
Material Science and Engineering, Texas A\&M University, College Station, TX 77845-3122, USA
}%
\author{Justin B. Haskins}
\affiliation{%
Artie McFerrin Department of Chemical Engineering, Texas A\&M University, College Station, TX 77845-3122, USA
}%
\author{Cem Sevik}
\affiliation{%
Artie McFerrin Department of Chemical Engineering \& Material Science and Engineering, Texas A\&M University, College Station, TX 77845-3122, USA \& Advanced Technologies Research Center, Anadolu University, Eskisehir 26470, Turkey}%
\author{Tahir {\c C}a{\u g}{\i}n}%
\email{tcagin@tamu.edu}
\affiliation{%
Artie McFerrin Department of Chemical Engineering \& Material Science and Engineering, Texas A\&M University, College Station, TX 77845-3122, USA
}%
\date{\today}

\begin{abstract}
Chemical and structural diversity present in hexagonal boron nitride ((\textit{h}-BN) and graphene hybrid nanostructures provide new avenues for tuning various properties for their technological applications. In this paper we investigate the variation of thermal conductivity ($\kappa$) of hybrid graphene/\textit{h}-BN nanostructures: stripe superlattices and BN (graphene) dots embedded in graphene (BN) are investigated using equilibrium molecular dynamics.  To simulate these systems, we have parameterized a Tersoff type interaction potential to reproduce the \textit{ab initio} energetics of the B-C and N-C bonds for studying the various interfaces that emerge in these hybrid nanostructures.
We demonstrate that both the details of the interface, including energetic stability and shape, as well as the spacing of the interfaces in the material exert strong control on the thermal conductivity of these systems.
For stripe superlattices, we find that zigzag configured interfaces produce a higher $\kappa$ in the direction parallel to the interface than the armchair configuration, while the perpendicular conductivity is less prone to the details of the interface and is limited by the $\kappa$ of \textit{h}-BN.
Additionally, the embedded dot structures, having mixed zigzag and armchair interfaces, affects the thermal transport properties more strongly than superlattices.
Though dot radius appears to have little effect on the magnitude of reduction, we find that dot concentration (50\% yielding the greatest reduction) and composition (embedded graphene dots showing larger reduction that \textit{h}-BN dot) have a significant effect.
\end{abstract}

\maketitle
\section{Introduction}

Following the isolation of single layer graphene,~\cite{NovoselovNat2007} studies on the electrical,~\cite{RevModPhys.81.109,Novoselov10222004,ISI:000248194700033,ISI:000233133500043,cresti} optical,~\cite{ISI:000274338800035,ISI:000256441100035} thermal,~\cite{Seol04092010, doi:10.1021/nl0731872,ISI:000279014300016, ISI:000277444900021, geim_acsnano,Balandin-NM-Rev} and mechanical~\cite{ISI:000257713900044, ISI:000253764000041}  properties of this low-dimensional material have revealed their potential for many technological applications.~\cite{doi:10.1021/nl070133j,PhysRevLett.100.206802, ISI:000244558400015,PhysRevLett.98.206805}
This in turn has triggered interest in isomorphs of graphene, namely \textit{h}-BN~\cite{PhysRevB.49.5081,ISI:000278888600004, ISI:000275858200036, ISI:000282727600056, ISI:000254669900075, ISI:000284990900046, ISI:000268652700007, ISI:000276905600060} and hybrid \textit{h}-BN/graphene structures.
Recently, fabrication of both random immersions of \textit{h}-BN in graphene~\cite{C1NR11387A,han:203112} and well-defined clusters of \textit{h}-BN in graphene with possible kinetically controllable domain sizes~\cite{ISI:000276953500024} has intensified this interest.
In particular, such hybrid systems have a considerable compositional and structural diversity that translates into greater freedom for tuning the physical properties.
Both experimental and density functional theory (DFT) studies have shown that the physical properties of these materials can be significantly modified by simply varying the relative amount of \textit{h}-BN to graphene.~\cite{doi:10.1021/nl2011142,lam:022101,qiu:064319}
For instance, Ci~\textit{et al.}~\cite{ISI:000276953500024} have experimentally shown that decreasing the relative amount of ~\textit{h}-BN to graphene increases the electrical conductivity, which has been supported by DFT studies where increasing BN concentration and cluster size results in band gap opening.~\cite{xu:073711,PhysRevB.84.125401}
It is recently shown that the details of the bonding at the \textit{h}-BN/graphene interface can change the type of intrinsic doping of the system.~\cite{PhysRevB.84.205412}
Just to name a few other examples of how this chemical and structural diversity in this low dimensional hybrid system enable control over magnetic properties; zigzag-edges in ribbons have been suggested to lead to ferromagnetic behavior~\cite{PhysRevLett.87.146803} while more complex interfaces, like those present in \textit{h}-BN clusters embedded in graphene, can be antiferromagnetic ~\cite{PhysRevB.84.075405}  may also be mentioned.

Thermal transport in graphene with embedded \textit{h}-BN quantum dots has been studied recently using real-space Kubo approach.~\cite{PhysRevB.84.205444}
This study has shown that the decreasing dot size decreases the phonon mean free path (MFP) of both in-plane and out-of-plane modes considerably.
However, limited variation in MFP has been observed by changing the dot concentration at the smallest dot sizes.
In another study, the effect of BN nanodots on the heat current in graphene nanoribbons has been investigated by using non-equilibrium Green's functions and nonequilibrium (direct method) molecular dynamics.~\cite{refId0}
The authors claimed that there is a linear inverse relationship between the number of atoms at the interface and the heat current.
Although these studies provide valuable insight about the thermal transport, the thermal conductivity of graphene/\textit{h}-BN nanostructured systems has not been investigated systematically by considering superlattices with different nano-morphologies.
The objective of this study is to investigate the influence of the chemical and structural diversity present in hexagonal boron nitride ((h-BN) and graphene hybrid nanostructures on thermal transport and test possible pathways for tuning the thermal conductivity of these low dimensional hybrid structures. In this paper, we investigate the variation of thermal conductivity of hybrid graphene/h-BN nanostructures in particular: 1) stripe superlattice geometries while varying geometric parameters and composition and 2) BN (graphene) dots embedded in graphene (BN) as a function of dot-diameter and composition. The theoretical findings aim at providing  basis for potential thermal management applications in miniaturized devices.

We have previously calculated the lattice thermal conductivities of nanotubes, graphene and \textit{h}-BN based nanostructures~\cite{Che_2000_2,JustinACS,Cem-BN-PRB,cem-bn-szotop,doi:10.1021/nl2029333} with considerable accuracy and compared the results with available experimental data.~\cite{Balandin-NM-Rev}
In this study, we implement an accurate model for C-B and C-N interactions by employing DFT calculations in addition to our previous \textit{h}-BN potential. Using these Tersoff interatomic potentials, we calculated the lattice thermal conductivity of several possible graphene/\textit{h}-BN hybrid structures.
The rest of the report is organized as follows: First, the model utilized to develop the potential and the calculation methods for thermal conductivity are described.
Then, the validity of our potentials for studying hybrid nanostructures is demonstrated.
This is followed by a detailed description of the considered hybrid nanostructures and a discussion of the effect of structure and composition on lattice thermal transport properties.

\section{Method}

Equilibrium molecular dynamics simulations can be utilized to obtain instantaneous heat current ($\bm{J}$) or energy moment ($\bm{R}$) as a function of time.
Subsequently, thermal conductivity, $\kappa$ can be evaluated by using either the heat current autocorrelation function (Green-Kubo method)~\cite{Green_1954,Kubo_1957,Zwanzig_1965} or mean square displacement of the energy moment (Einstein relation)~\cite{Zwanzig_1965} as discussed in detail in our earlier studies.~\cite{JustinACS,Cem-BN-PRB,cem-bn-szotop,Alper,JustinNanotech}
Here, the thermal conductivity is evaluated from the Einstein relation (the mean square displacement of energy moment, named \emph{h}MSD) as given by~\cite{Weitz_1989}

\begin{equation}
\frac{\big\langle[R_\mu(t)-R_\mu(0)]^2\big\rangle}{2V k_BT^2} = \kappa_{\mu\mu} [t+\tau(e^{-t/\tau}-1)].
\label{eq:Reinstein}
\end{equation}

Here, $V$ is the volume, $T$ is the temperature and $k_B$ is the Boltzmann constant.
The energy moment through direction $\mu$ is defined by $R_\mu$.
The right hand side of Eq.~\ref{eq:Reinstein} represents a linear change in Einstein relation for the time ($t$) much larger than the decay time ($\tau$).
The long-time behavior corresponds to diffusive regime in transport of heat.
For short-times, on the other hand, the average energy propagation is ballistic and results in a non-linear relation between $\kappa$ and \emph{h}MSD.
Given the time, a bulk system assumes a diffusive behavior at elevated temperatures and thus we are more interested in this regime.
Computationally, we eliminate the non-linear portion of the relationship by discarding the first 100 ps of \emph{h}MSD then fit the rest to a linear function, i.e., \emph{h}MSD $=2V k_BT^2\kappa_{\mu\mu}t$, in order to obtain thermal conductivity.

\begin{figure}[!h]
\includegraphics[width=14.0cm]{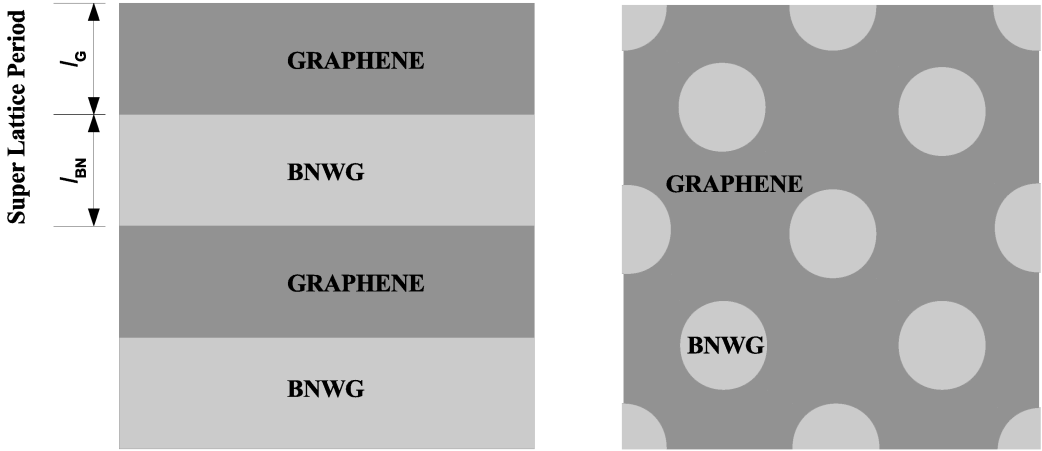}
\caption{\label{fig1}(Color online) The hybrid structures structures considered in this work \emph{viz.}: stripe superlattices and dots embedded in sheet matrix.}
\end{figure}

In this study, we investigate the thermal conductivity of graphene/\textit{h}-BN superlattices in the form of stripes and dots/``anti"dots, see Fig.~\ref{fig1}.
The stripe superlattices are discussed in two general categories. In the first case, equal periods ($l_{\mathrm{G}}$ = $l_{\mathrm{BN}}$), and in the second unequal periods ($l_{\mathrm{G}}$ $\neq$ $l_{\mathrm{BN}}$) of graphene and \textit{h}-BN stripes are simulated.
The stripes of graphene and \textit{h}-BN sublattices are connected \textit{via} two different orientations namely, resulting in a zigzag or an armchair interface.
For all structures, approximately 60$\times$60~$nm^2$ periodic domains are considered.
Previously, we showed that such large systems are required for the proper convergence of thermal conductivity in equilibrium MD calculations of ribbon like systems.~\cite{JustinACS}
For the equal period simulations, in each orientation, five different period thicknesses ranging from $\sim$1.25 to $\sim$30 $nm$ are constructed.
The atomistic details of these systems are given in Table~\ref{table1} in the Appendix.
For the unequal period simulations, again five different configurations are created for the armchair and zigzag interface systems where the thicknesses of BN sublattices change from $\sim$3 to $\sim$57 $nm$ and the sum of $l_{\mathrm{BN}}$ and $l_{\mathrm{G}}$ is set to $\sim$60 $nm$, see Table~\ref{table2} in the Appendix for details.

As a second type of nanostructure, dots of~\textit{h}-BN are embedded in graphene with a close-packed arrangement as shown in Fig.\ref{fig1}.
We select three different radii (4.95 \AA, 12.38 \AA, and 24.76 \AA) for these ordered dots.
Ordered graphene dots in~\textit{h}-BN, so called anti-dots, are created with radius of 12.38 \AA. Random configuration of antidots are also considered with radii of 6.19 and 12.38 \AA.
The details of these structures are provided in Table~\ref{table3} in the Appendix.

Molecular dynamics simulations are performed in the microcanonical (NVE) ensemble with a time step of 1.0~fs to conserve energy and a simulation length of 5~ns to obtain an acceptable ensemble average of \textit{h}MSD.
Each data point for $\kappa$ is therefore obtained by averaging the results of a minimum of five distinct simulations with different initial velocity distributions.
The error in $\kappa$ value is calculated from the standard deviation of these independent calculations.
The volumes of the two-dimensional structures are defined as $lw\Delta$, where $w$ and $l$ are the width and the length of simulated structures, and $\Delta$ (0.335 $nm$) is the mean Van der Waals thickness for \textit{h}-BN and graphene.
Finally, we did not consider isotopic disorder explicitly in thermal conductivity calculations.
Instead, a single mass of natural abundance is used for all elements.

We have previously developed a Tersoff-type potential for \textit{h}-BN systems.~\cite{Cem-BN-PRB}
Also, a Tersoff parametrization for graphene is given by Lindsay and Broido.~\cite{PhysRevB.81.205441}
Both potentials have been optimized to reproduce the DFT phonon dispersions for their respective material, necessary for ensuring accurate lattice thermal conductivities.
In order to simulate the interfaces, one needs to further develop interaction potentials for all possible element pairs coupling at the interface.
We have used DFT calculations to generate the data needed for the interfaces.
These calculations have been performed with Vienna \textit{ab initio} simulation package (VASP)~\cite{vasp1,vasp2} which is based on density functional theory.
Projector augmented wave (PAW)~\cite{PAW1, PAW2} pseudo potential formalism was imposed with Perdew-Burke-Ernzerhof (PBE)~\cite{GGA1} form of generalized gradient approximations (GGA).
Using DFT energetics to condition empirical potentials has been previously motivated for both pure graphene and \textit{h}-BN.
The PAW-PBE formalism, in particular, produces accurate structures for our systems of interest, with the calculated lattice parameters for graphene and \textit{h}-BN being 2.45\AA{} and 2.51\AA{}, respectively.
Long horizontal strips of the structures in Fig.~\ref{fig2} are used in periodic-boundary conditions in order to avoid spurious interface-interface interactions.
Depending on the basic repeating unit of the given structures, in-plane dimensions of 29.95~\AA{} $\times$2.47~\AA{} (structures 2 and 3), 30.22~\AA{} $\times$2.49~\AA{} (structures 1 and 4), or 24.8~\AA{} $\times$4.30~\AA{} (structure 5)  were used with 2$\times$16$\times$1, 2$\times$16$\times$1, and 2$\times$10$\times$1 Monkhorst-Pack $k$ point grids, respectively.
400~eV is selected for the plane wave energy cut-off to achieve the energy convergence.

\section{Results and Discussions}

\subsection{Optimization of C-BN Parameters}

\begin{figure}[!h]
\includegraphics[width=12.0cm]{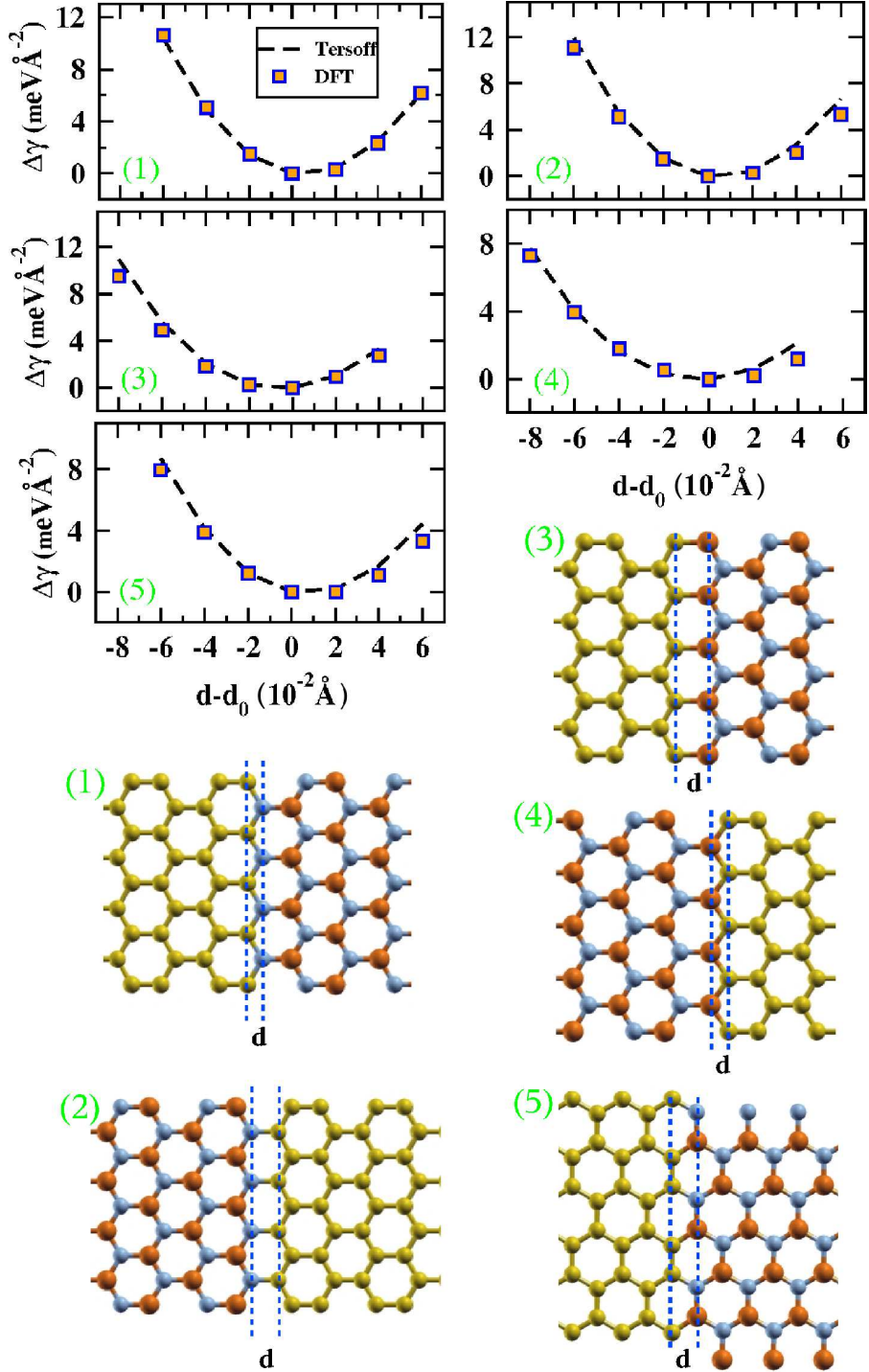}
\caption{\label{fig2}(Color online) The change in the total energy, as given by both the Tersoff potential and DFT, per interface area as a function of interface separation, d. The corresponding structures for the interfaces are given with ball-and-stick representations (C = yellow, N = small blue, B = big red atom).}
\end{figure}

As pointed put in the previous section, reliable potentials for C-C and B-N interactions have appeared in the literature.
To simulate the hybrid structures of interest, we must then only define the interactions between B-C and N-C.
Since the structure and vibrational spectrum of \textit{h}-BN and graphene are similar, we opt to employ the mixing rules and fitting procedure put forth by Tersoff for Si-Ge and Si-C,~\cite{PhysRevB.39.5566} which approximates the parameters as a mixture of the existing BN and C parameters modified by two arbitrary values, $\chi_{\mathrm{B-C}}$ and $\chi_{\mathrm{N-C}}$.
These parameters adjust the contribution from the attractive term to the potential.
We have obtained $\chi_{\mathrm{B-C}}$ and $\chi_{\mathrm{N-C}}$  by imposing the requirement on the potential to reproduce DFT energetics of all probable \textit{h}-BN/graphene interfaces shown in Fig.~\ref{fig2}.
In these graphs, $\Delta\gamma$ is the change in total energy per interface area (width $\times$ Van der Waals thickness) as the interface separation, d, changes from the equilibrium value, d$_0$, under the condition that the bond lengths in the graphene and \textit{h}-BN regions are held fixed.
The corresponding interfaces are also shown in Fig.~\ref{fig2}.
Note that the interface separation parameter, d, accounts for both bond length and angle variations.
Parameter fitting is accomplished by minimizing the differences between the DFT and the force field derived $\Delta\gamma$ values for each displacement for each structure simultaneously by updating the force field parameters using a genetic algorithm.
The fitted parameters for $\chi_{B-C}$ and $\chi_{N-C}$ (0.886777 and 1.013636 respectively) along with the parameters obtained from the mixing rule have produced MD energies in good agreement with DFT results.
Moreover, the error in the calculated equilibrium B-C and N-C bond lengths of all structures is no larger than 1.5\%.
A full list of parameters along with the description of interaction potential function and the mixing rules are given in Table~\ref{table4} of the Appendix.

\subsection{Stripe Superlattices with Equal Periods}

\begin{figure}[!h]
\includegraphics[width=14.0cm]{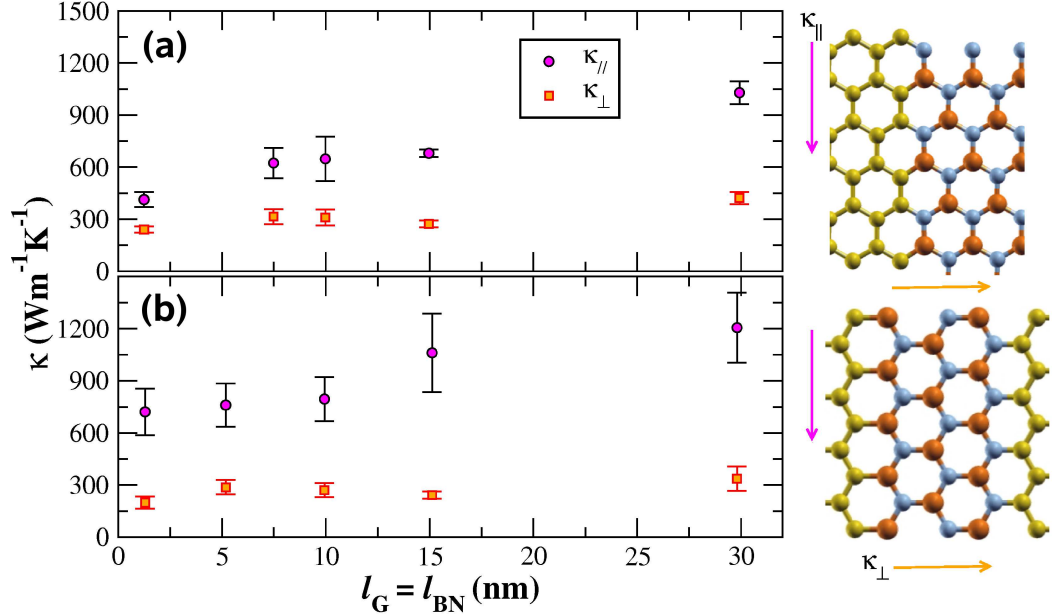}
\caption{\label{fig3}(Color online) The thermal transport coefficients parallel and perpendicular to the two different graphene/\textit{h}-BN interfaces are shown in (a) and (b). The period lengths of both graphene and \textit{h}-BN are constrained to be equal. The atomistic details for the calculated structures are given in Table~\ref{table1} in Appendix.}
\end{figure}

In all striped superlattice structures, we calculated the lattice thermal conductivities parallel, $\kappa_{\shortparallel}$, and perpendicular, $\kappa_{\perp}$, to the superlattice orientation.
The chosen interfaces are shown in Fig.~\ref{fig3} and~\ref{fig4} with the associated thermal conductivity values.
The ball-and-stick structure in Fig.~\ref{fig3}a and in Fig.~\ref{fig4}a is the same interface given in Fig.~\ref{fig2} as structure 5, essentially one armchair ribbon connected to the other two in a symmetrical fashion though forming B-C and N-C bonds.
Whereas the structure represented as Fig.~\ref{fig3}b and Fig.~\ref{fig4}b can be thought of as one zigzag ribbon connected to two others on one side by B-C bonds and on the other by N-C bonds. These interfaces correspond to structures 2 and 3 in Fig.~\ref{fig2}.
The effective stiffness at the interface, obtained by fitting the $\Delta\gamma$ to a quadratic function, shows that the C-N bond is stronger than C-B bond.
This is expected considering that both interactions are mainly covalent and as more electrons are involved in the bonding, the strength of the bond increases.

Fig.~\ref{fig3} shows how the thermal conductivity of the aforementioned superlattice interfaces behave when the periods, constrained by $l_{\mathrm{G}}$ = $l_{\mathrm{BN}}$, are varied. The transport coefficients in the parallel direction, however, behave differently, depending on the type of interface. The superlattice with armchair interface has smaller thermal conductivity compared to the one with zigzag interface in the studied period range. As the period thickness increases, the stripe structures appear to become less sensitive to interface effects on  parallel thermal conductivity for both interfaces approaching ~1050-1200 $W/mK$. This is close to the midpoint of the thermal conductivity values of pristine \textit{h}-BN, 450 $W/mK$, and graphene, 2300 $W/mK$. This behavior agrees quantitatively with what is expected from treating the striped structure as the combination of two independent nanoribbons.
Previously, it was shown that zigzag ribbons have better thermal transport properties than armchair ribbons at small widths because of the latter having higher atomic line density on the edge.~\cite{JustinACS,Cem-BN-PRB} Thicker ribbons have more transport channels and difference in scattering behavior at the edges become less significant. Thus, it is sensible for striped structures combined through zigzag interfaces to have larger transport coefficients in smaller periods. 
Another apparent observation is that the thermal transport coefficients perpendicular to the different interfaces behave similarly, gradually increasing from 200-250 $W/mK$ at $l = 1.5~nm$ to 350-400 $W/mK$ at $l = 30~nm$. The perpendicular thermal transport is strongly controlled by the lower thermal conductivity component (\textit{h}-BN) and the interface phonon scattering even at a 30~nm thickness. If one assumes that the periods of the stripes are longer than the phonon mean free path and the boundary resistance is negligible, then $\kappa_{\perp}$ of the stripe system of equal periods is bounded by $2(\kappa_{\mathrm{graphene}}\times\kappa_{\mathrm{\textit{h}-BN}})/(\kappa_{\mathrm{graphene}}+\kappa_{\mathrm{\textit{h}-BN}})$.
For the calculated superlattices this equation give 752.7 $W/mK$.
The actual physics of the simulated systems, on the other hand, will not resemble to the idealized picture.
First, the system has a finite thermal boundary resistance that depends on the acoustic mismatch of the stripes and the intrinsic properties of the boundary. The effect of boundary structure on $\kappa_{\perp}$ is less pronounced when the results from Fig.~\ref{fig3} a) and b) compared, and it is almost independent for zigzag and armchair interfaces. 
Second, some of the systems have period lengths of only few nanometers which is very short compared to the MFP of the relevant phonons.
Thermal conductivity perpendicular to the interface increases slowly as the period size grows; however, the ideal value will not be reached because of the limiting effect of thermal boundary resistance, which will be present even in systems with period sizes longer than the characteristic MFP.

\subsection{Stripe Superlattices with Unequal Periods}

\begin{figure}[!h]
\includegraphics[width=14.0cm]{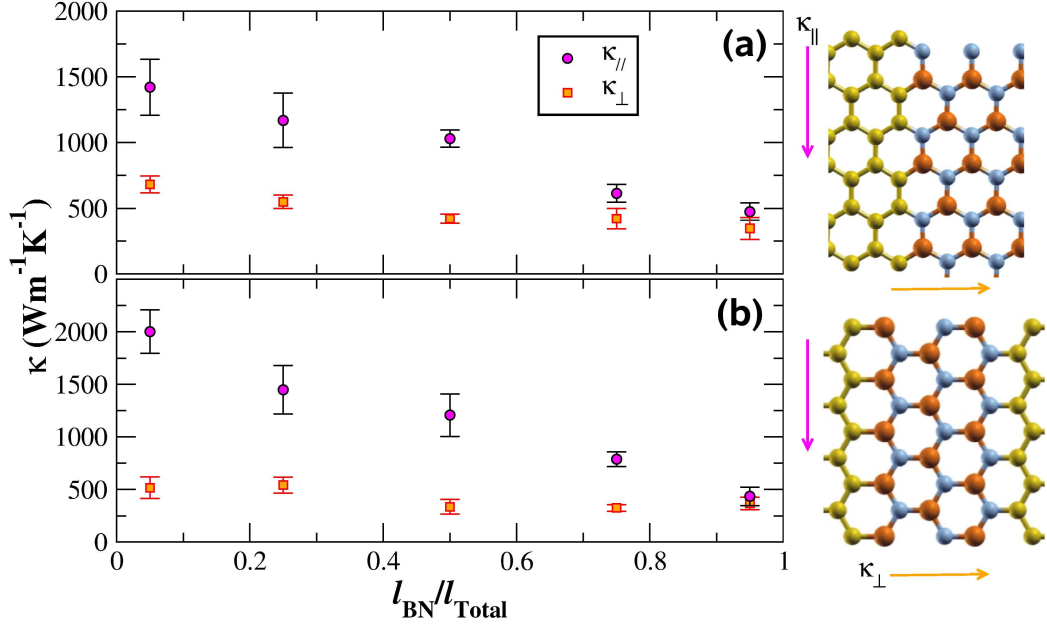}
\caption{\label{fig4}(Color online) The thermal transport coefficients parallel and perpendicular to the two different graphene/\textit{h}-BN interfaces are shown in (a) and (b). The sum of period lengths of graphene and \textit{h}-BN are constrained to be 60 nm. The atomistic details for the calculated structures are given in Table~\ref{table2} in Appendix.}
\end{figure}

Using the same interfaces, we remove the constraint of equal size periods and only require the sum of $l_{\mathrm{BN}}$ and $l_{\mathrm{G}}$ to be 60 $nm$.
We note here that the variation of the period lengths also enables us to see the influence of concentration.
When \textit{h}-BN has a small concentration (or a small period), the parallel component of thermal transport increases toward the limiting value of graphene as seen in Fig.~\ref{fig4}.
On the other hand, the perpendicular component does not exceed 700 $W/mK$.
Again, the zigzag interfaces have higher parallel thermal transport coefficients (35\% larger) than the armchair interfaces in almost all configurations.
When the period of BN is small, the reduction in $\kappa_{\perp}$ from the pristine graphene value is mainly due to interfacial phonon scattering; systems with larger $l_{BN}$ drive $\kappa_{\perp}$ toward the pure BN values but are still limited by the influence of interfacial scattering. The effect of atomic bonding at the interface on conduction is most clearly seen when $l_{BN}/l_{Total}$=0.05. Conductivity perpendicular to the boundary in armchair interfaced sample is noticeably higher than zigzag sample. This is most probably caused by enhanced scattering from alternating types of interface bonding in zigzag boundaries. 

\subsection{Dot and Anti-dot Superlattices}

\begin{figure}[!h]
\includegraphics[width=14.0cm]{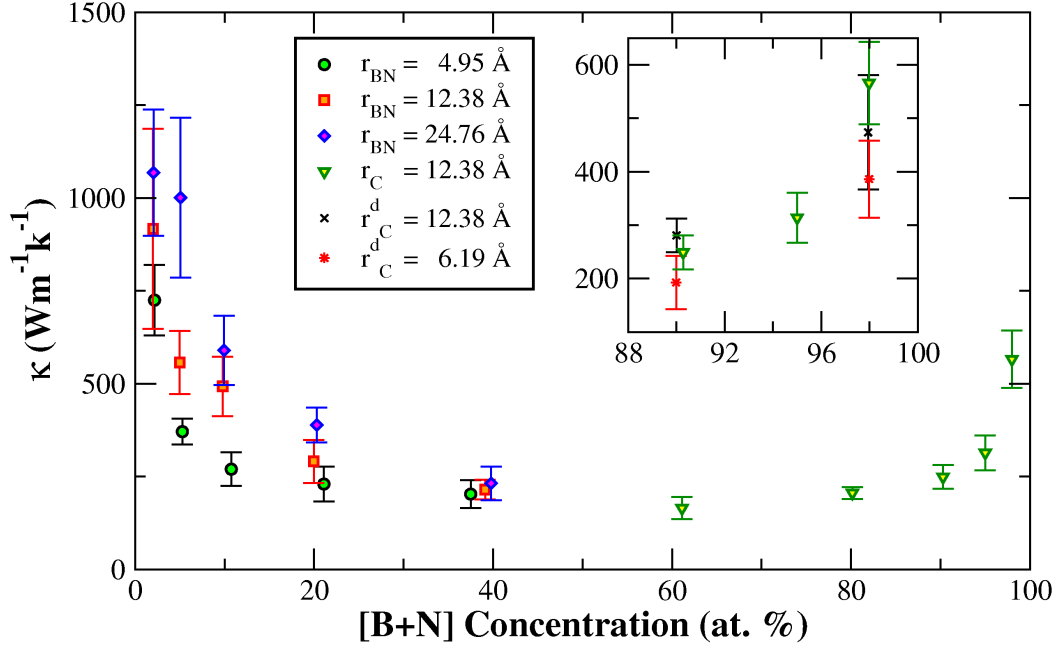}
\caption{\label{fig5}(Color online) The thermal transport properties of graphene with embedded \textit{h}-BN dots and \textit{h}-BN with embedded graphene. Three different radii, 4.95 \AA, 12.38 \AA, and 24.76 \AA~are used for \textit{h}-BN dots. The number of dots in the systems are varied such that the BN concentration ranges from ~2-98\%. Two different radii, 6.19 \AA, 12.38 \AA~are employed for graphene dots. The superscript ``d'' indicates the disordered dot arrangement. BN concentration on the horizontal axis is calculated as the percent ratio of the total number of boron and nitrogen atoms to the total number of atoms. The inset graph has the same axis units as the outer graph. The atomistic details for these systems are given in Table~\ref{table3} in Appendix.}
\end{figure}

We now turn to the investigation of the thermal conductivity of ordered and random distributions of~\textit{h}-BN dots embedded in graphene.
Fig.~\ref{fig5} shows the influence of dot size and concentration on the $\kappa$.
From Fig.~\ref{fig5} we see that larger dot sizes lead to higher thermal transport coefficients.
At the lowest BN concentration (2\%) the system with the largest dot has a 20\% larger transport coefficient than the other sizes.
This could be understood by the fact that larger dots have a smaller boundary to bulk ratio at the same concentration.
As more dots are introduced, this interface effect is suppressed and the $\kappa$ of all systems converge to 250 $W/mK$ at 40\% \textit{h}-BN.
Interestingly, this large concentration limit is similar to the perpendicular conductivity of stripe superlattices with periods similar to the diameter of the dots, see Fig.~\ref{fig3}.
It is likely that at large concentrations the \textit{h}-BN dots can isotropically limit the thermal transport in the same manner that the stripes limit the transport perpendicular to the boundary.

In addition to ordered BN dots, we have modeled ordered and random distributions of graphene dots in \textit{h}-BN.
The thermal conductivity values of these systems are also presented in Fig.~\ref{fig5}.
A decreasing behavior in thermal conductivity is also observed in these systems as the number of graphene dots increases.
It is surprising to see graphene, as the higher $\kappa$ component, does not enhance the thermal conductivity of ~\textit{h}-BN.
This can be attributed to the relatively small size of the dots and the large \textit{h}-BN/graphene interface to area ratio, leading to interfacial phonon scattering events dominating $\kappa$.
At the lowest C concentration, the ordered dot system has higher thermal conductivity than the bulk value of \textit{h}-BN.
It is not clear whether this is an actual physical phenomena or an averaging problem since the error bars are large enough to include the bulk value.
In creating the random dot configurations, we maintain the mean dot separation similar to the one in the ordered configurations with same concentration.
For each concentration, the initial conditions of the simulations are not only varied by atom velocities but the also the distribution of the dots.
The thermal conductivities of the structures, having ordered and random dots, are not significantly different for the same dot sizes and concentrations (see the inset of Fig.~\ref{fig5}).
Again, the smaller dots lead to lower $\kappa$ when the concentration of C is kept constant.

\section{Summary and Concluding Remarks}

We have characterized the lattice thermal transport properties of hybrid graphene and \textit{h}-BN structures: graphene-white graphene stripes and dot/antidot superlattices.
The $\kappa_{\perp}$ of striped nanostructures with large periods is limited by the less conductive component, \textit{h}-BN.
The parallel transport, on the other hand, attains a value close to the average of the two components.
As the periods of the stripes are reduced, interface scattering effects become more prevalent with zigzag interfaces resulting in higher $\kappa$ than the armchair interfaces.
The thermal conductivity of the dot systems can be tailored by both dot diameter and concentration.
Small dot concentration and large dot diameter leading to larger conductivities.
Moreover, the transport properties of nanosystems with high dot concentrations are independent of size, approaching the $\kappa_{\perp}$ of the small period striped superlattices.

\section{Acknowledgements}
We acknowledge support from NSF (DMR 0844082) to the International Institute of
Materials for Energy Conversion at Texas A\&M University as well as AFRL. Parts
of computations were carried out by the Laboratory of Computational Engineering
of Nanomaterials, supported by ARO, ONR and DOE grants. We would also like to
thank the Supercomputing Center of Texas A\&M University for a generous time
allocation for this project. CS acknowledges the support from The
Scientific and Technological Research Council of Turkey (TUBITAK) to his
research at Anadolu University.

\clearpage

\section{Appendix}

Structural details of stripe and dot superlattices are given in Table~\ref{table1}, Table~\ref{table2} and Table~\ref{table3}.

\begin{scriptsize}\begin{table}[!h]
\caption{\label{table1} Simulation details for the stripe superlattices where graphene and \textit{h}-BN have equal period thicknesses. Thermal conductivities of these structures are given in Fig.~\ref{fig3}. The total size of the systems are given by Length$_{arm}$ and Length$_{zig}$ where the subscripts define whether the length is measured along the armchair or the zigzag configuration.}
\begin{ruledtabular}
\begin{tabular}{lccccc}
Boundary $\backslash$ Period ($nm$) & \# of B & \# of N & \# of C & Length$_{arm}$ ($nm$)& Length$_{zig}$ ($nm$)\\\hline
Armchair $\backslash$ $l=1.246854$ & 33600 & 33600  & 67200 & 60.46925 &  59.84897 \\
Armchair $\backslash$ $l=7.481121$ & 33600 & 33600 & 67200 & 60.46925 & 59.84897  \\
Armchair $\backslash$ $l=9.974828$ & 33600 & 33600 & 67200 & 60.46925 & 59.84897 \\
Armchair $\backslash$ $l=14.9622$ & 33600  & 33600 & 67200 & 60.46925 &  59.84897 \\
Armchair $\backslash$ $l=29.92449$ & 33600  & 33600 & 67200 & 60.46925 & 59.84897 \\
Zigzag $\backslash$ $l=1.2957698$  & 32982 & 32982 & 65964 & 59.60541 & 59.59960 \\
Zigzag $\backslash$ $l=5.1830792$  & 34560 & 34560 & 69120 & 62.19695 & 59.84897 \\
Zigzag $\backslash$ $l=9.934235$ & 32982 & 32982 & 65964 & 59.60541 & 59.59960  \\
Zigzag $\backslash$ $l=15.1173$ & 33600 & 33600 & 67200 & 60.46925 & 59.84897  \\
Zigzag $\backslash$ $l=30.234625$ & 33600 & 33600 & 67200 & 60.46925 & 59.84897 \\\hline
\end{tabular}
\end{ruledtabular}
\end{table}\end{scriptsize}

\newpage

\begin{scriptsize}\begin{table}[!h]
\caption{\label{table2} Simulation details for the stripe superlattices where graphene and \textit{h}-BN have different period thicknesses. Thermal conductivities of these structures are given in Fig.~\ref{fig4}. The total size of the systems are given by Length$_{arm}$ and Length$_{zig}$ where the subscripts define whether the length is measured along the armchair or the zigzag configuration.}
\begin{ruledtabular}
\begin{tabular}{lccccc}
Boundary $\backslash$ $l_{\mathrm{BN}}/l_{\mathrm{total}}$  & \# of B & \# of N & \# of C & Length$_{arm}$ ($nm$)& Length$_{zig}$ ($nm$)\\\hline
Armchair $\backslash$ 0.05 & 3360 & 3360 & 127680 & 60.46925 & 59.84897 \\
Armchair $\backslash$ 0.25 & 16800 & 16801 & 100800 & 60.46925 & 59.84897 \\
Armchair $\backslash$ 0.50 & 33600 & 33600 & 67200 & 60.46925 & 59.84897 \\
Armchair $\backslash$ 0.75 & 50400 & 50400 & 33600 & 60.46925 & 59.84897 \\
Armchair $\backslash$ 0.95 & 63840 & 63840 & 6720 & 60.46925 & 59.84897 \\
Zigzag $\backslash$ 0.05 & 3360 & 3360 & 127680 & 60.46925 & 59.84897 \\
Zigzag $\backslash$ 0.25 & 16800 & 16800 & 100800 & 60.46925 & 59.84897 \\
Zigzag $\backslash$ 0.50 & 33600 & 33600 & 67200 & 60.46925 & 59.84897 \\
Zigzag $\backslash$ 0.75 & 50400 & 50400 & 33600 & 60.46925 & 59.84897 \\
Zigzag $\backslash$ 0.95 & 63840 & 63840 & 6720 & 60.46925 & 59.84897 \\\hline
\end{tabular}
\end{ruledtabular}
\end{table}\end{scriptsize}

\newpage

\begin{scriptsize}\begin{table}[!h]
\caption{\label{table3} Simulation details for graphene with embedded \textit{h}-BN dots and \textit{h}-BN with embedded graphene anti-dots. Thermal conductivities of these structures are given in Fig.~\ref{fig5}.}
\begin{ruledtabular}
\begin{tabular}{lccccc}
Radius ($nm$) & \# of B & \# of N & \# of C & Length$_{X}$ ($nm$)& Length$_{Y}$ ($nm$)\\\hline
& 6000 & 4800 & 18000 & 29.71200 & 25.73130\\
& 3840 & 3072 & 25856 & 31.69280 & 27.44680\\
r$_{\mathrm{BN}}=$ 0.495& 2016 & 1728 & 31104 & 32.68320 & 28.30446\\
& 960 & 768 & 31040 & 31.69280 & 27.44676\\
& 540 & 432 & 44028 & 37.14000 & 32.16417\\\hline
& 12384 & 12960 & 39456 & 44.56800 & 38.59704\\
& 5504 & 5760 & 45184 & 41.59680 & 36.02388\\
r$_{\mathrm{BN}}=$ 1.238& 3096 & 3240 & 58464 & 44.56800 & 38.59701\\
& 1376 & 1440 & 53632 & 41.59680 & 36.02388\\
& 348 & 360 & 34140 & 32.68320 & 28.30448\\\hline
& 13032 & 12744 & 39024 & 44.56800 & 38.59701\\
& 5792 & 5664 & 44992 & 41.59680 & 36.02388\\
r$_{\mathrm{BN}}=$ 2.476& 1448 & 1416 & 25936 & 29.71200 & 25.73135\\
& 1448 & 1416 & 53584 & 41.59680 & 36.02389\\
& 1448 & 1416 & 136528 & 65.36640& 56.60896\\\hline
& 14000 & 13500 & 17500 & 37.14000 &32.16420 \\
& 22784 & 22464 & 11200 & 41.59680 & 36.02388 \\
r$_{\mathrm{C}}=$ 1.238& 29340 & 29160 & 6300 & 44.56800 & 38.59701 \\
& 26864 & 26784 & 2800 & 41.59680 & 36.02388 \\
& 68320 & 68256 & 2816 & 65.36640 & 56.60896 \\\hline
r$^{\mathrm{d}}_{\mathrm{C}}=$ 1.238 & 26750 & 26820 & 5950 & 39.88552 & 40.15493 \\
& 29150 & 29145 & 1225 & 39.87575 & 40.14509 \\\hline
r$^{\mathrm{d}}_{\mathrm{C}}=$ 0.619 & 26795 & 26765 & 5960 & 39.89808  & 40.16757 \\
& 29160 & 29160 & 1200 & 39.88407 & 40.15347 \\\hline
\end{tabular}
\end{ruledtabular}
\end{table}\end{scriptsize}

\newpage

The potential used in this study is developed by Tersoff.~\cite{PhysRevB.39.5566}

\begin{eqnarray}
V_{ij} &=& f_{C}(r_{ij})\left[f_{R}(r_{ij})+b_{ij}f_{A}(r_{ij})\right]\nonumber\\
f_{C}(r_{ij}) &=& \left\{\begin{array}{rcl}
 1 & : &r_{ij} < R_{ij} \\
\frac{1}{2}+\frac{1}{2}\cos\left(\pi\frac{r_{ij}-R_{ij}}{S_{ij}-R_{ij}}\right) & :
&R_{ij} < r_{ij} < S_{ij} \\ 0 & : &r_{ij} > S_{ij}
\end{array}\right. \nonumber\\
f_{R}(r_{ij}) &=& A_{ij} \exp\left(-\lambda^{I}_{ij}r_{ij}\right) \nonumber\\
f_{A}(r_{ij}) &=& -B^{'}_{ij} \exp\left(-\lambda^{II}_{ij}r_{ij}\right) ,\;\;\;\;\;\; B^{'}_{ij} = B_{ij}\chi_{ij} \nonumber\\
b_{ij} &=&\left(1+\beta_{i}^{n_{i}}\zeta_{ij}^{n_{i}}\right)^{-\frac{1}{2n_{i}}} \nonumber\\
\zeta_{ij} &=& \sum_{k\neq i,j} f_{C}(r_{ik}) g(\theta_{ijk}) \nonumber \\
g(\theta_{ijk}) &=& \left(1+\frac{c_{i}^{2}}{d_{i}^{2}}-\frac{c_{i}^{2}}{\left[d_{i}^{2}+(\cos\theta_{ijk}-h_{i})^{2}\right]}\right)\nonumber
\end{eqnarray}

In this description the lower indices $i$, $j$ and $k$ mark the atoms where $i$-$j$ bond is modified by a third atom $k$. The potential parameters and their corresponding values are given in Table~\ref{table4}. The parameter $\chi_{ij}$ was used as a fitting parameter in our study. For the mixing of parameters, the geometric mean is calculated for the multiplier parameters and arithmetic mean is calculated for the exponential parameters. These rules are given below.

\begin{eqnarray}
\lambda^{I}_{ij} &=& \left(\lambda^{I}_{i}+\lambda^{I}_{j}\right)/2,\;\;\;\;  \lambda^{II}_{ij} \;=\;
\left(\lambda^{II}_{i}+\lambda^{II}_{j}\right)/2,\;\;\;\; A_{ij} \;=\;\left(A_{i}A_{j}\right)^{(1/2)}\nonumber\\
B_{ij} &=& \left(B_{i}B_{j}\right)^{(1/2)},\;\;\;\; R_{ij} \;=\;\left(R_{i}R_{j}\right)^{(1/2)},\;\;\;\; S_{ij} \;=\;\left(S_{i}S_{j}\right)^{(1/2)}\nonumber
\end{eqnarray}

It should be mentioned that $\chi_{ij}$ modifies $B_{ij}$ which is obtained as a result of the mixing procedure. Here, we also note that the developed potential is not parameterized to represent N-N or B-B interactions as can be seen from Table~\ref{table4}.

\newpage
\begin{scriptsize}\begin{table}[!h]
\caption{\label{table4} The parameters of Tersoff potential optimized for C-BN interactions. The atom X represent the bond modifying element where all parameters are exactly the same whether it is C, B or N.}
\begin{ruledtabular}
\begin{tabular}{lccccccc}
Parameters& C B X & C C X & C N X & B C X & B N X & N B X & N C X\\\hline
$A$ (eV)&1386.78&1393.6 &1386.78 &1386.78 & 1380.0  & 1380.0 & 1386.78 \\
$B^{'}$ (eV)&339.06891&430.0  &387.575152 &339.068910 & 340.0 & 340.0 & 387.575152  \\
$\lambda^{I}$ (\AA$^{-1}$)&3.5279   &3.4879 &3.5279 & 3.5279 & 3.568 & 3.568 & 3.5279 \\
$\lambda^{II}$ (\AA$^{-1}$)&2.2054   &2.2119 &2.2054 & 2.2054 & 2.199 & 2.199 & 2.2054 \\
$n$&0.72751&0.72751 &0.72751&0.72751&0.72751&0.72751&0.72751 \\
$\beta$ (10$^{-7}$)&1.5724   &1.5724  &1.5724 & 1.25724 & 1.25724 & 1.25724 & 1.25724 \\
$c$&38049&38049 &38049 &25000 & 25000 & 25000 & 25000 \\
$d$&4.3484&4.3484 &4.3484 &4.3484 &4.3484 &4.3484 &4.3484 \\
$h$&-0.93&-0.93 &-0.93 & -0.89 & -0.89 & -0.89 & -0.89 \\
$R$ (\AA)&1.85&1.80 &1.85 &1.85 & 1.90 & 1.90 & 1.85 \\
$S$ (\AA)&2.05&2.10 &2.05 &2.05 & 2.00 & 2.00 & 2.05 \\\hline
\end{tabular}
\end{ruledtabular}
\end{table}\end{scriptsize}

\end{document}